\documentclass[a4paper,UKenglish,cleveref, autoref, thm-restate]{lipics-v2021}

\usepackage{soul}
\usepackage{xcolor} 

\title{Towards Competence-Based Management for Open Source
Software Projects} 
\usepackage{booktabs}
\usepackage{tikz}
\titlerunning{Towards Competence-Based Management for Open Source
Software Projects}
\author{Sabahat Younas}{Independent Researcher, Pakistan}{sabahatyounas7@gmail.com}{}{}
\author{Márcia Moraes}{Colorado State University, Fort Collins, United States}{marcia.moraes@colostate.edu}{}{}
\author{Fabio Santos}{Colorado State University, Fort Collins, United States}{fabiomarcosdev@gmail.com}{}{}

\authorrunning{S. Younas, M. Moraes, and F. Santos}

\Copyright{Sabahat Younas, Márcia Moraes, and Fabio Santos}


\relatedversion{} 

\nolinenumbers 

\EventEditors{Robert Feldt, Maria Paasivaara, Daniel Mendez, Stefan Wagner, and Marvin Mu\~{n}oz Bar\'{o}n}
\EventNoEds{5}
\EventLongTitle{20th International Symposium on Empirical Software Engineering and Measurement (ESEM 2026)}
\EventShortTitle{ESEM 2026}
\EventAcronym{ESEM}
\EventYear{2026}
\EventDate{October 8--9, 2026}
\EventLocation{Munich, Germany}
\EventLogo{}
\SeriesVolume{394}
\ArticleNo{64}

\begin{document}

\maketitle

\begin{abstract}
  Contributors to Open Source Software (OSS) projects are vital to maintaining the health of both communities and projects. However, the number of projects experiencing core contributors' disengagement has increased to the point that it risks the projects' survival. 
  Finding new contributors to replace the workforce is challenging and time-intensive due to several factors, including a lack of precise knowledge about potential candidates' competences, which may need to be confirmed through interviews and exams. 
  Previous studies provided indications of the contributor's competences, although they lack depth in understanding competence levels, which can result in poor knowledge about the contributor's capabilities. To address this gap, we assess contributors' competence by collecting code metrics related to source code from contributions. We also propose a competence model able to predict the competence level required to solve tasks. By properly assessing contributors' competences, we can identify and train candidates to replace core contributors. Our replication package, including code, data, and documentation, is available at 
\url{https://doi.org/10.5281/zenodo.21605122}  
\end{abstract}

\section{Introduction}
Open Source Software (OSS) exhibits increasing project counts, file counts, and churn, while the number of joiners remains stable or declines in most projects \cite{zhou2017scalability}. Many of the projects depend on a few core contributors, with numbers approaching 1 or 2 in about 70 of the most popular GitHub repositories \cite{avelino2016novel}, and projects tend to die when abandoned by maintainers due to vital technical debt \cite{avelino2019abandonment}. 

Competence-based management is a key factor in Human Resources (HR) selection, development, planning, and performance management \cite{rodriguez2002developing}; therefore, knowing the core contributors' competences can help recruit, onboard, and train new members. 
Competence is commonly described as a combination of attitude, skills, and knowledge \cite{assyne2022essential}. However, it remains challenging to consistently identify, represent, and assess human competences \cite{calhau2021towards}. Previous studies have demonstrated the importance of revealing the capabilities of the workforce and the existing tasks in OSS context. Liang et al. \cite{liang2022understanding} used surveys to investigate skills that support the development of OSS projects. Dey et al. \cite{dey2021representation} showed that contributors are more likely to join projects aligned with their competence and have pull requests accepted. The onboarding of newcomers to OSS projects and the selection of tasks using proxies for skills were also the subjects of studies \cite{santos2023tag, santos2021can}. However, previous studies are limited in depth since they ignored: differences between programming languages \cite{dey2021representation}, security vulnerability \cite{dey2021representation}, and the intricacy of competence levels \cite{dey2021representation, santos2023tag, santos2021can}. The depth limit affects the performance of predictive models used to suggest issues for contributors~\cite{carter2025skillscope, vargovich2023givemelabeledissues}, thereby mitigating a core contributor shortage and supporting expert recruitment.  



To close the gap of competence depth, we present a case study using source code data to evaluate contributors' competence in OSS, including levels mined from 5,793 unique functions. Our approach combines Abstract Syntax Tree (AST)-based competence extraction, hierarchical competence modeling, code complexity metrics (13 Lizard metrics), and machine learning models to build contributors' profiles and predict competence levels from pull requests. The profiles can later be used to generate recommendations for training or search for contributors to replace workforce as needed. Considering that, we are interested in answering the following research questions (RQ1): To what extent can we predict the competences to solve issues? (RQ2): How can contributor competences be automatically identified and profiled? 

By answering these RQs, this work aims to contribute with (1) a novel hierarchical approach for competence profiling that aggregates function-level complexity metrics to competences based on usage patterns, (2) preliminary validation on a manually verified set of pull requests from JabRef demonstrating up to 90\% accuracy in predicting competence levels for well-represented technical domains using a reduced, leakage-aware set of code complexity features, (3) empirical evidence that code metrics and structure are stronger predictors of competence requirements, and (4) a scalable, automated method for profiling competences to later use for training, planning task allocation and replacement when core maintainers disengage, therefore contributing to the competence management.

\subsection{Related Work}

The approach combines language-aware competence extraction with structural evidence and proficiency levels to address limitations in prior research. Unlike Dey et al. \cite{dey2021representation}, who use vector representations from API changes, this approach preserves programming-language semantics and analyzes API and library usage through AST-level structure. It shifts the focus from task/issue skill signals \cite{santos2023tag, santos2021can} to code contributions and models levels of competence using code complexity and structural metrics. Hannan et al. \cite{DBLP:conf/esem/HannanRRS25} used Large Language Models (LLMs) to identify competence and cosine similarity to match tasks and coders; however, the method did not reveal the level of competence. Additionally, unlike Liang et al.'s \cite{liang2022towards, liang2022understanding} broad \mbox{human-centered} model, this study operationalizes competence in a measurable and automatable way from repository artifacts, aiding in scalable profiling and decision support for training and replacement when core maintainers disengage.

\section{Methodology}

\subsection{Case Study Design}

We designed a case study to explore archival data to create a competence model and build competence profiles looking for evidence of their utility in measuring OSS workforce competence ~\cite{runeson2009guidelines}.

\begin{figure}[ht]
\vspace{0.3cm}
\centering
\includegraphics[width=\linewidth]{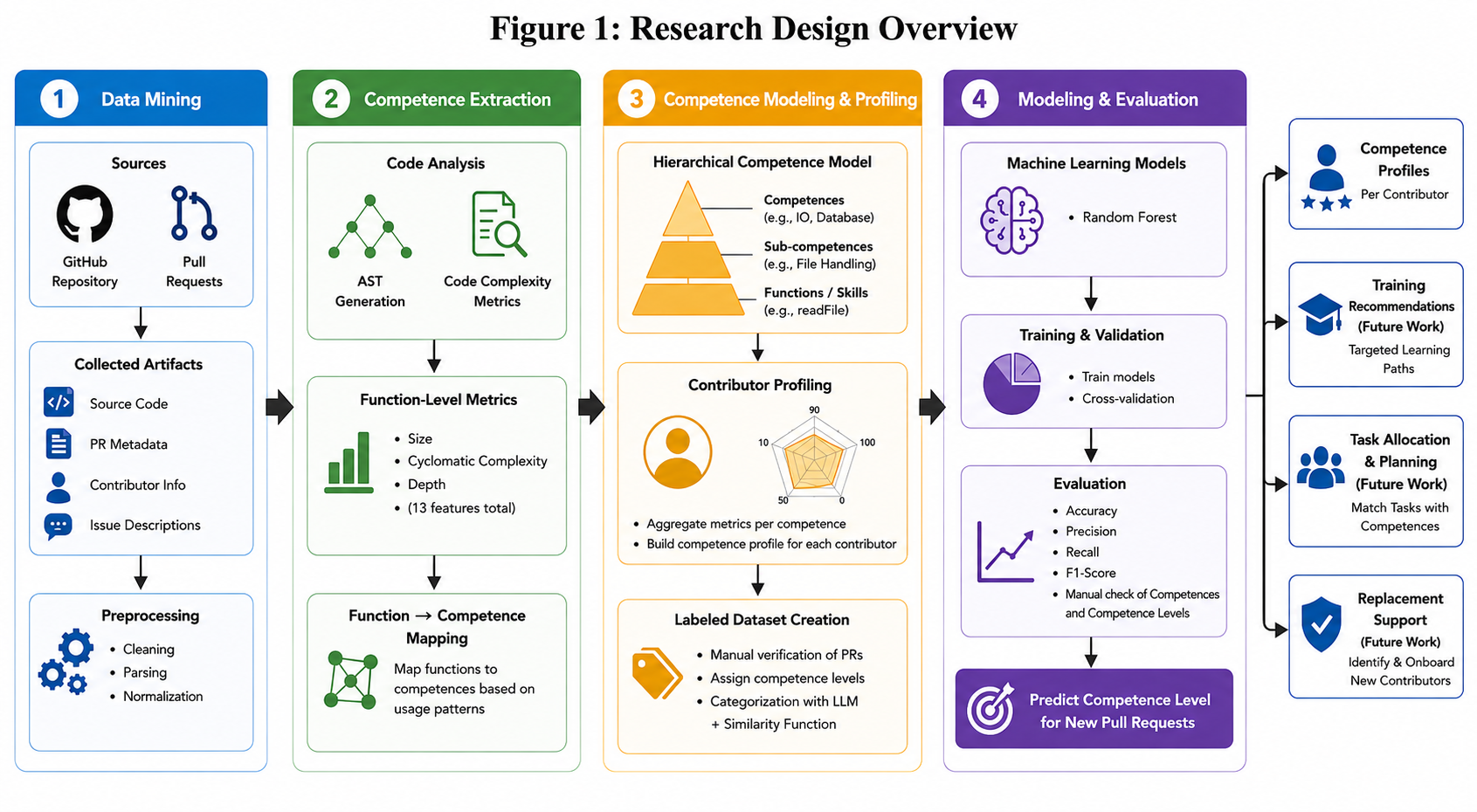}
\vspace{0.2cm}




\caption{Overview of the proposed competence profiling workflow.}
\label{fig:workflow}
\end{figure}

Our approach follows a systematic workflow composed of four main components: (1) data mining, (2) competence and sub-competence extraction, (3) complexity analysis and level assignment, and (4) model training (Figure \ref{fig:workflow}).

In the data collection section (Section \ref{sec:collection}), we detail the data gathering process including the justification for the data source, which was the OSS project targeted for this study and how our data was organized to find evidence of the competences based on the source code and their respective AST with the support of a taxonomy and a Large Language Model (LLM) for categorization. Next, we gathered static metrics from the source code to derive the levels using a composite score (Section \ref{sec:complexityanalysis}). Finally, we built a prediction model (Section \ref{sec:model}) to verify to what extent the proposed competence management can address the proposed motivation.

\subsection{Data Collection}\label{sec:collection}

Data collection comprised the mining process (Section \ref{sec:datamining}) and competence extraction (Section \ref{sec:competenceextraction}) 

\subsubsection{Data Mining}
\label{sec:datamining}

We collect data from the JabRef GitHub repository, a mature OSS Java project used as a case study in this work. We selected JabRef based on its project maturity and the breadth of technical competences represented in its codebase (e.g., database management, UI, network communication, parsing), which is important for testing a competence taxonomy spanning multiple domains. It is a popular OSS Java-based reference management tool with over 4,000 stars on GitHub, more than 900 contributors, 4,500 closed issues, and a history of research collaboration, where maintainers and researchers find solutions for OSS projects. It also has a complex codebase spanning multiple technical competences including database management, user interface design, network communication, and bibliographic parsing. The broad range of competences is important for testing our predictive model. Finally, one author contributed to JabRef, enabling concrete evaluation of the competences. For this case study, we limited the dataset size to allow a rigorous verification of the AST results and categorization. The dataset consists of 5,793 unique functions. The functions collected encompass 85 PRs where two authors (one JabRef contributor and one software engineer) could evaluate the competence categorization.

From the repository, we extract all source files modified in merged PRs together with PR metadata, including author information, timestamps, changed files, and contribution statistics. These data are used as input for competence extraction, complexity analysis, and model training. 

We filter source files by removing non-code files and focusing on substantial code contributions which contain at least one parsable source code file from which AST structures and Lizard complexity metrics can be extracted. These filtered source files feed into our competence extraction pipeline, while the PR metadata is used for both competence extraction and model training phases. We sent the source code from the 5,793 unique functions to the competence-level extraction pipeline.

\subsubsection{Competence Extraction}\label{sec:competenceextraction}

For each selected source file, we perform the following steps:

\textbf{Abstract Syntax Tree (AST) Generation:} We parse each source file into its AST representation, capturing the complete structural and syntactic information of the code.

\textbf{Extract APIs and Function Names:} From the generated ASTs, we traverse the tree structure to identify and extract all API calls, library imports, and function/method definitions used in the code.

\textbf{Competence Taxonomy:} Using the extracted APIs and function names, we build a hierarchical competence taxonomy ~\cite{carter2025skillscope} which contains 217 competences organized in two layers (31 competences in the first layer and 186 sub-competences in the second layer). This taxonomy organizes programming constructs into meaningful concepts (e.g., ``Database - Backup,'' ``IO - File Management,'' and ``UI - Event Handler,'' where Database is the first layer, or the competence and Backup, the second layer, or the sub-competence). 

\textbf{Competence Classification:} We classify each API and function into concepts from the taxonomy generated in the previous phase, producing classifications for each source file. This output maps each of the 5,793 unique functions declared in the source files to the competence taxonomy, capturing competence and sub-competence.

\textbf{LLM-based Categorization:} We used the zero-shot approach similar to SkillScope \cite{carter2025skillscope}, but employing LLaMA-3.1-8B-Instant \cite{grok} via the Groq API to categorize APIs and functions into our competence taxonomy. We chose this approach because \mbox{LLaMA-3.1-8B-Instant} is an open-access model that offers a competitive cost-benefit. The LLM receives only taxonomy definitions without example classifications, using a near-zero temperature (temperature=0.1) to limit output variability while avoiding the degenerate or repetitive completions that strict determinism (temperature=0) can sometimes produce~\cite{laimouche2025comparison}. To mitigate hallucinations, we implement post-processing by ingesting the LLM output into spaCy \cite{spacy} semantic similarity, to ensure outputs align with valid taxonomy entries. Complete prompt templates are available in our replication package. Two authors evaluated a sample of 361 (from the 5,793 unique functions, using a 95\% confidence level and 5\% margin of error) functions to ensure correct classification.

The extracted PR metadata are combined with the competence and sub-competence categories defined in the taxonomy, using one-hot encoding, to create comprehensive training datasets that link contributors to their demonstrated competence. 

\subsection{Data Analysis}
\label{sec:complexityanalysis}
Using the Lizard static analysis tool \cite{yin2019lizard}, we extract comprehensive function-level code complexity metrics from each modified file in the pull request. Rather than relying solely on cyclomatic complexity, we employ a composite scoring approach that combines multiple complementary metrics to provide a more robust assessment of code difficulty and required competence.

Lizard computes cyclomatic complexity (CCN) by counting linearly independent paths through the code: each conditional statement (if, else if), loop (for, while, do-while), case in a switch statement, logical operator in conditions (and, or), and exception handler (catch) increments the CCN by one. Additionally, Lizard provides non-comment lines of code (NLOC), token count, parameter count, and function length metrics.

To avoid the limitations of single-metric approaches, we calculate a composite complexity score using weighted contributions from multiple metrics:

\vspace{-1em}
\begin{equation}
\text{Composite Score} = \sum_i w_i \times \frac{M_i}{T_i}
\end{equation} \label{eq:compositescore}
\vspace{-1em}

Where $M_i$ represents each metric value, $T_i$ is its normalization threshold, and $w_i$ is its weight. The metrics and their weights are: \textbf{CCN Metrics (30\% total):} average CCN (10\%), maximum function CCN (10\%), and CCN density (10\%). \textbf{Size Metrics (50\% total):} total NLOC (10\%), average NLOC per function (10\%), average  parameters (10\%), average function length (10\%), total tokens (5\%), and average tokens per function (5\%). \textbf{Structural Metrics (20\% total):} number of functions (5\%), number of files (5\%), number of competences (5\%), and number of sub-competences (5\%). 


Each metric is normalized to a 0-1 scale using empirically-derived thresholds based on JabRef data analysis. We use different normalization strategies depending on the metric distribution. 

For metrics with highly skewed distributions and extreme outliers, such as NLOC and token counts, we use the 95th percentile as the normalization threshold to reduce the influence of unusually large pull requests. For example, NLOC is normalized by 100 (95th percentile observed: 52) and tokens by 200 (95th percentile observed: 435).

For metrics with more stable ranges and bounded behavior, such as CCN and function counts, we use empirically observed maximum reference values. CCN values are normalized by 30 (maximum observed: 78), while function counts are normalized by 100. This normalization strategy ensures that metrics with different scales contribute proportionally to the composite score while limiting distortion caused by extreme values.

We map composite scores to five competence levels: scores $\leq$ 0.25 indicate simple code (Level 1), 0.25 $<$ score $\leq$ 0.45 suggests moderate complexity (Level 2), 0.45 $<$ score $\leq$ 0.65 indicates complex code (Level 3), 0.65 $<$ score $\leq$ 0.85 represent very complex code (Level 4), and score $>$ 0.85 denotes highly complex code requiring Level 5 competence.

This composite approach addresses the limitation that cyclomatic complexity alone may not fully capture all dimensions of code difficulty. For example, a function with moderate CCN but extremely high token count (indicating dense logic) or a function with low CCN but spanning many lines (indicating structural complexity) would both be appropriately weighted in the final assessment. The weights were tuned using a Grid Search technique aiming to increase the accuracy of the prediction model and prioritize decision-making complexity 
while accounting for code size and project structure. The number of complexity levels and thresholds were inspired by the work of Liang et al.~\cite{liang2022towards}. Previous studies have similarly used static analysis metrics obtained from source code to investigate code violations, cognitive complexity, and the security and quality characteristics of LLM-generated code~\cite{haindl2024does, kharma2026security, santa2025llm}, supporting the use of such metrics as a meaningful signal for competence assessment in this study.

Similar function-level combinations of static code metrics (NLOC, cyclomatic 
complexity, token count, and parameter count) alongside Term Frequency-Inverse Document Frequency (TF-IDF) features 
have been used effectively for lightweight software engineering classification 
tasks, such as vulnerability triage in C/C++ codebases~\cite{chiu2025lightweight}, 
supporting the use of this metric family as an established, non-arbitrary 
foundation for our composite scoring approach.

To determine whether the surrogate model's predictive performance differs significantly from a naive baseline, rather than diverging only by chance, we apply McNemar's test~\cite{mcnemar1947note}, a non-parametric test for paired nominal data that is well suited to comparing two classifiers evaluated on the same set of samples, as recommended by Dietterich~\cite{dietterich1998approximate} for the statistical comparison of supervised learning algorithms.

\subsection{Model Training}\label{sec:model}
We implemented a hierarchical complexity extraction and prediction system using the Lizard static analysis tool to measure code complexity at multiple levels of granularity. Our system extracts 13 distinct complexity metrics and aggregates them hierarchically from the function level; of these, a reduced set of 6 structural metrics (Table~\ref{tab:ml-features}) is used as input to the surrogate competence level model.

\begin{table}[ht]
\tiny
\centering
\vspace{0.3cm}
\caption{Features for Competence Level Prediction}
\label{tab:ml-features}
\small
\begin{tabular}{lp{4.5cm}p{5cm}}
\toprule
\textbf{Feature} & \textbf{Description} & \textbf{Competence Signal} \\
\midrule
Total NLOC & Sum of non-comment code lines & Contribution size \\
Total Tokens & Sum of lexical tokens & Lexical density \\
Num Functions & Count of modified functions & Breadth of change \\
Num Files & Count of modified files & Cross-file impact \\
Num Competences & Count of technical domains & Domain diversity \\
Num Sub-competences & Count of sub-domains & Fine-grained breadth \\
\bottomrule
\end{tabular}
\vspace{0.2cm}
\end{table}



\textbf{Hierarchical Aggregation:} 
We aggregate function-level metrics following a three-tier hierarchy: \textbf{Function Level:} Individual function metrics from Lizard. \textbf{Sub-competence Level:} Functions are grouped by API usage patterns (e.g., "Database → Query Execution", "User Interface → Layout Design"). For each sub-competence, we calculate: total CCN, total NLOC, total tokens, number of functions, average CCN, average NLOC per function, average tokens per function, and maximum function CCN. \textbf{Competence Level:} Sub-competences are aggregated into parent competences, and metrics are the sum of all sub-competence metrics within that competence.

\textbf{Composite Complexity Scoring:} We calculate a weighted composite score from all 13 metrics to assign competence levels. The composite score combines the metrics using equation (1) and competence levels (Section \ref{sec:competenceextraction}).



This multi-metric approach provides a more robust assessment than single-metric thresholds, as it captures multiple dimensions of code complexity. 

\textbf{Machine Learning Models:}  We adopt the one-vs-all strategy~\cite{mirza2013one} to deal with the large number of categories in the multi-label problem aggravated by the data imbalance. One-vs-all strategy also enables the study to scale to more layers, potentially leading to an extreme multi-label problem in the future~\cite{wydmuch2018no}. 

We train two complementary models:

\textit{Competence Classification:} Predicts which technical areas are involved in a pull request using issue and linked PR text (title and description) processed through TF-IDF vectorization following previous research indicating best performance with lexical approaches instead of semantic for noisy text \cite{DBLP:conf/esem/HannanRRS25}. 
This multilabel model identifies which technical areas are involved.

\textit{Competence Level Prediction Model:} Predicts the required competence level (1-5 scale) using both textual features 
and a surrogate model with six code complexity metrics~\cite{guyon2003introduction}. The option for the surrogate model is twofold: it deals with the data leakage of predicting the levels using the same features used to obtain them, and enables the model to use only essential data available when the PR is submitted. This comparative feature-reduction approach follows standard practice in feature selection~\cite{guyon2003introduction, hastie2009elements}.
This multiclass model determines \textit{who} (at what competence level) should work on the contribution.

The models work in tandem to provide comprehensive competence profiling: the  competence model identifies technical scope, while the competence level model quantifies required experience. The predictive models demonstrate that the proposed metrics capture signals aligned with the requirements for identifying competence levels. We ran the models using Random Forest (RF) which reported competitive results in Software Engineering research~\cite{carter2025skillscope, santos2023tag, santos2021can,  vargovich2023givemelabeledissues}.  



\section{Results}


In this section, we present the results of the study. 

\subsection{Dataset Characteristics}

We successfully extracted complexity metrics from 19,005 function-level records (a function counted once for each commit in which it was modified), corresponding to 5,793 functions uniquely identified by name and file path across the dataset. From the 85 pull requests analyzed, we discarded 21 PRs that did not declare any source code and contained documentation or configuration-only updates. In total, this yielded 64 PRs that contained substantive code modifications with at least one extracted function and measurable complexity metrics. Table~\ref{tab:dataset} summarizes the dataset characteristics.

\begin{table}[ht]
\vspace{-5pt}
\tiny
\centering
\vspace{0.3cm}
\caption{Dataset Summary Statistics (n=64)}
\label{tab:dataset}
\begin{tabular}{lr}
\toprule
\textbf{Metric} & \textbf{Value} \\
\midrule
Total PRs Analyzed & 85 \\
PRs with Code Changes & 64 (75.3\%) \\
Empty PRs (docs/config) & 21 (24.7\%) \\
\midrule
Avg Functions per PR & 296.9 \\
Avg NLOC per PR & 4,411  \\
Avg Tokens per PR & 32,598  \\
\midrule
Min Functions & 3 \\
Max Functions & 3,196 \\
Min Sub-competences & 1 \\
Max Sub-competences & 33 \\
\bottomrule
\end{tabular}
\vspace{0.2cm}
\end{table}

Note that the 296.9 average functions per PR (Table~\ref{tab:dataset}) is computed from 
the 19,005 total function-level records, not the 5,793 uniquely identified functions, 
since a function modified across multiple commits is counted once per commit.

We organize technical knowledge using a two-level hierarchy composed of competences and sub-competences. Competences represent broad technical areas (e.g., User Interface, Network, Database), while sub-competences capture more specific activities within them (e.g., Layout Design, Connection Management, Query Execution). Each pull request may involve one or multiple competences and sub-competences depending on the APIs and functions identified during AST analysis and classification.

The dataset exhibits substantial variance in complexity, ranging from simple, single sub-competence changes (e.g., PR \#34 with 8 functions in one sub-competence) to large architectural modifications (e.g., PR \#80 with 2,156 functions across 33 sub-competences and 15 competences). This diversity enables robust model training across different competence levels.

McNemar's test confirmed the surrogate model's improvement over a majority-class dummy baseline was not due to chance: across 5,793 unique functions, the surrogate was correct where the dummy was wrong 981 times, versus 369 in the reverse, yielding a significant McNemar statistic of 276.53 ($p < 0.0001$).

\subsection{Complexity Metrics Distribution}

Table~\ref{tab:complexity} presents the distribution of extracted complexity metrics across all 64 PRs with code changes.

\begin{table}[ht]
\vspace{0.2cm}
\tiny
\centering
\caption{Complexity Metrics Distribution}
\label{tab:complexity}
\begin{tabular}{lrrr}
\toprule
\textbf{Metric} & \textbf{Mean} & \textbf{Min} & \textbf{Max} \\
\midrule
Cyclomatic Complexity & 3.32 & 1.00 & 78.00  \\
Lines of Code (NLOC) & 4,411 & 24 & 50,671 \\
Token Count & 32,598 & 175 & 367,105 \\
Functions per PR & 296.9  & 3 & 3,196 \\
Sub-competences per PR & 7.2 & 1 & 33 \\
Competences per PR & 5.1 & 1 & 15 \\

\bottomrule
\end{tabular}
\vspace{0.2cm}
\end{table}

The mean cyclomatic complexity of 3.32 indicates that most code in JabRef is moderately complex, falling into the Level 1 - Level 2 range. However, the maximum CCN of 78.00 suggests that some contributions require Level 5 competence. The wide range in NLOC (24 to 50,671) and tokens (175 to 367,105) reflects the diversity of contribution sizes, from small, low-line-count changes to major feature additions.





\subsection{Model Performance}

Despite the relatively small dataset, both models achieved strong performance. 
Table~\ref{tab:model-performance} summarizes the results.

\begin{table}[ht]
\tiny
\vspace{0.3cm}
\centering
\caption{Model Performance Metrics}
\label{tab:model-performance}
\begin{tabular}{llr}
\toprule
\textbf{Model} & \textbf{Metric} & \textbf{Value} \\
\midrule
\multirow{3}{*}{Competence Classification} & Accuracy & 92.91\% \\
 & F1-Score & 75.80\% \\
 & Training Samples (functions) & 5,793  \\
\midrule
\multirow{4}{*}{Surrogate Model (6 features)}
 & Test Accuracy & 69.23\% \\
 & F1-Score (weighted) & 67.18\% \\
 & CV Accuracy (3-fold) & 62.41\% \\
 & Training Samples (functions) & 5,793  \\
 \midrule
 \multirow{2}{*}{Dummy Baseline (Competence Level, most-frequent)} & Accuracy & 38.46\% \\
 & F1-Score (weighted) & 21.37\% \\
 & Training Samples (functions) & 5,793  \\
\bottomrule
\end{tabular}
\vspace{0.2cm}
\end{table}


The Competence Classification model achieved 92.91\% accuracy in identifying which technical competences are involved in a PR. For competence level prediction, we adopted a surrogate model using six structural features to mitigate the risk of target leakage inherent in predicting complexity levels from complexity-derived metrics themselves. The surrogate competence level model (69.23\% accuracy, 67.18\% F1) substantially outperforms the strongest naive dummy baseline (38.46\% accuracy, 21.37\% F1), supporting that it captures genuine predictive signal from code complexity metrics rather than exploiting class imbalance alone. 

The competence level distribution across our 64 PR-level training samples was Level 1 (20.3\%), Level 2 (30.8\%), Level 3 (40.6\%), Level 4 (6.2\%), and Level 5 (2\%), with contributions concentrated in the moderate-complexity range (Levels 2--3) and few reaching the highest complexity tiers. Note this reflects individual PR labels prior to aggregation; Section~\ref{contributors} reports a markedly different distribution once competence is aggregated to the contributor level using a maximum rule (Table~\ref{tab:contributors}).

\subsection{Per-Domain Performance Analysis}

To understand how the competence level prediction model performs across different technical domains, we analyzed prediction accuracy for each of the competences identified in our dataset. Table~\ref{tab:surrogate-domain-performance} presents the results, sorted by F1-Score.

\begin{table}[ht]
\tiny
\vspace{0.3cm}
\centering
\caption{Competence Level Prediction Accuracy (Surrogate Model), 
by Technical Domain}
\label{tab:surrogate-domain-performance}
\begin{tabular}{lrrrrrr}
\toprule
\textbf{Domain} & \textbf{Accuracy} & \textbf{Precision} & \textbf{Recall} & \textbf{F1-Score} & \textbf{Samples} & \textbf{Avg Level} \\
\midrule
Databases & 90.00 & 91.67 & 90.00 & 89.45 & 10 & 2.30 \\
Error Handling & 85.71 & 89.29 & 85.71 & 84.35 & 7 & 2.14 \\
Operating System & 75.00 & 76.94 & 75.00 & 75.43 & 12 & 2.50 \\
Computer Graphics & 70.00 & 70.09 & 70.00 & 69.97 & 30 & 2.50 \\
Input-Output & 68.97 & 73.78 & 68.97 & 69.27 & 29 & 2.17 \\
Event Handling & 65.52 & 65.80 & 65.52 & 65.34 & 29 & 2.45 \\
Network & 63.64 & 69.26 & 63.64 & 64.69 & 11 & 2.55 \\
User Interface & 63.41 & 65.83 & 63.41 & 63.62 & 41 & 2.56 \\
Language & 57.14 & 65.00 & 57.14 & 60.32 & 14 & 2.71 \\
Data Structure & 56.25 & 56.35 & 56.25 & 56.00 & 32 & 2.41 \\
\bottomrule
\end{tabular}
\vspace{0.3cm}
\end{table}


The surrogate model achieved up to 90.00\% prediction accuracy in the Databases domain, demonstrating its effectiveness in predicting contributor competence in specialized technical areas. Performance decreased in broader and more heterogeneous domains such as Language (57.14\%) and Data Structure (56.25\%), where the wider variety of development tasks and complexity patterns makes competence-level prediction more challenging. Overall, the results indicate that domain specialization enables more accurate prediction, whereas heterogeneous domains introduce greater classification uncertainty. Label cardinality is 7.1094 and label density is 0.0846, confirming the multilabel nature of the problem and the integrity of the learning process~\cite{blanco2019multi, MultilabelBook}. 

\subsection{Feature Importance Analysis}

Feature importance analysis revealed that code complexity metrics substantially 
outperform textual features. Table~\ref{tab:surrogate-features} shows the top features.


\begin{table}[ht]
\tiny
\vspace{0.2cm}
\centering
\caption{Surrogate Model Feature Importance (6 Features)}
\label{tab:surrogate-features}
\begin{tabular}{lr}
\toprule
\textbf{Feature} & \textbf{Importance} \\
\midrule
Total Tokens & 0.2725 \\
Total NLOC & 0.2324 \\
Num Files & 0.1561 \\
Num Functions & 0.1428 \\
Num Competences & 0.1058 \\
Num Sub-competences & 0.0904 \\
\bottomrule
\end{tabular}
\vspace{0.2cm}
\end{table}

Total token count emerged as the strongest predictor (importance: 0.2725), followed by total NLOC (0.2324) and number of files touched (0.1561). Structural breadth metrics (number of competences and sub-competences) contributed less, suggesting that raw contribution size and lexical density carry more signal for level prediction than the breadth of technical areas touched.

\textbf{RQ1 Answer:} The competence level can be predicted with up to 90\% accuracy in the best-performing domain (Databases) using a reduced, leakage-aware surrogate model; the most influential predictors are total token count and total NLOC, indicating that contribution size and lexical density are stronger signals than technical breadth alone.





\subsection{Contributor Profiling}
\label{contributors}

Beyond analyzing individual pull requests, our approach enables the construction of comprehensive contributor profiles by aggregating competence across all contributions. For each contributor, we apply a maximum aggregation rule: the highest observed level in each sub-competence is retained in the contributor profile. 
This reflects the principle that when managers want to search for a contributor for a possible allocation, they can start by searching for the highest-skilled contributor, and then investigate the individual competence/sub-competence levels to reveal more details about the contributor's competences that demonstrate capability at a certain level. 

\textbf{Contributor Analysis Results:} From 5,793 unique functions, we identified 16 contributors with varying competence levels. 
The contributor-level distribution below reflects, per contributor, the maximum 
level reached across their individual sub-competence scores, not the PR-level distribution 
reported earlier. Because each contributor spans many sub-competences (avg. 20.5, Table~\ref{tab:top-contributors}), a high level in even one area is enough 
to register, so contributor-level Level 5 rates run higher than PR-level rates.
Table~\ref{tab:contributors} presents the distribution of contributors by their maximum competence level achieved.

\begin{table}[ht]
\tiny
\vspace{0.2cm}
\centering
\caption{Contributor Distribution by Competence Level}
\label{tab:contributors}
\begin{tabular}{lrr}
\toprule
\textbf{Max Competence Level} & \textbf{Count} & \textbf{Percentage} \\
\midrule
Level 5  & 3 & 18.75\% \\
Level 4  & 2 & 12.5\% \\
Level 3  & 7 & 43.75\% \\
Level 2  & 4 & 25.0\% \\
\bottomrule
\end{tabular}
\vspace{0.2cm}
\end{table}

The distribution reveals that 56.25\% of contributors (9 of 16) demonstrate Level 3 or Level 4 capabilities in at least one technical area, suggesting that a substantial proportion of contributors possess intermediate to advanced technical capabilities. This expertise distribution can help project maintainers identify contributors with similar technical capabilities when planning contributor replacement or assigning new tasks.

\textbf{Competence Breadth Analysis:} Contributors demonstrated competence across an average of 20.5 competence/sub-competence combinations (SD = 10.8, range = 6-43). Table~\ref{tab:top-contributors} shows the most versatile contributors.

\begin{table}[ht]
\tiny
\vspace{0.2cm}
\centering
\caption{Top Contributors by Competence Breadth}
\label{tab:top-contributors}
\begin{tabular}{lrrr}
\toprule
\textbf{ID} & \textbf{Areas} & \textbf{Max Lvl} & 
\textbf{PRs} \\
\midrule
d & 43 & 5 & 
4 \\
c & 39 & 5 & 
3 \\
0 & 35 & 5 & 
4 \\
b & 28 & 4 & 
5 \\
\bottomrule
\end{tabular}
\vspace{0.2cm}
\end{table}

Contributor "d" demonstrated the broadest competence, achieving Level 5 proficiency in four competences while maintaining capabilities across 43 distinct sub-competence areas. This breadth-and-depth combination makes such contributors particularly difficult to replace, as finding candidates with comparable versatility requires matching not just peak competence but also the range of secondary competences.

\textbf{Competence Profile Example:} To illustrate the granularity of our profiles, Table~\ref{tab:contributor-profile} shows the competence distribution for contributor "a", who achieved Level 4 status across multiple competences.

\begin{table}[ht]
\tiny
\vspace{0.2cm}
\centering
\caption{Competence Profile for Contributor "a"}
\label{tab:contributor-profile}
\begin{tabular}{lllr}
\toprule
\textbf{Level} & \textbf{Competence} & \textbf{Sub-competence} & \textbf{PRs} \\
\midrule
4 & Computer Graphics & Image Rendering & 4 \\
4 & Data Structure & Linear Structures & 5 \\
4 & Event Handling & User Interactions & 5 \\
4 & User Interface & Layout Design & 5 \\
\midrule
3 & Error Handling & Exception Handling & 1 \\
3 & Input-Output & Data Reading & 3 \\
3 & Network & Protocol Implementation & 1 \\
3 & Operating System & Memory Management & 1 \\
\multicolumn{3}{l}{...and 19 additional areas at Levels 1-2} & \\
\bottomrule
\end{tabular}
\vspace{0.2cm}
\end{table}

This profile reveals that contributor "a" achieved Level 4 mastery in four core areas, Level 3 competence in four additional areas, and maintained working proficiency across 
19 other technical competences at Levels 1--2. Such detailed profiles enable  precise matching: if a project needs to replace this contributor, the search can prioritize candidates with demonstrated Level 4 capabilities in at least Computer Graphics and 
User Interface competences, rather than simply seeking "experienced Java developers." 

\textbf{RQ2 Answer:} The proposed method identified contributor profiles aggregated from historical contributions revealing competences, sub-competences and levels. 

\section{Discussion}

\subsection{Implications for OSS Sustainability}

Competence modeling has been explored by the industry to understand how to improve performance with qualified personnel~\cite{calhau2024core}. Our study mines data from OSS repositories aiming to build profiles to benefit OSS communities by utilizing contributor's competences to optimize task allocation towards populating an ontology to anchor the concepts~\cite{Calhau2023zooming, calhau2021towards} for automatic matching~\cite{faria2013agreementmakerlight} at the next stage of this research.

Our approach has several implications for OSS project sustainability. First, by enabling automated candidate search at scale, projects can proactively build contributor pipelines before disengagement occurs. Second, the deep competence profiling allows for more accurate matching, reducing onboarding friction and increasing the likelihood of successful integration. Third, maintainers can identify competence gaps in their current team and train or recruit strategically to improve project resilience.


\subsection{Limitations}

External Validity: Case studies are limited in generalization on their own. Internal Validity: Metrics derived from static code analyzers may not always accurately reflect real-world code complexity; nevertheless, they are widely used tools in industry to give feedback to coders. LLM categorization cannot ensure determinism. However, we mitigated this through a low sampling temperature (0.1) to reduce output variability, combined it with a similarity metric to filter out hallucinations, and we manually analyzed 361 samples (functions) to confirm the results.
Construct validity: Competence management is a broad concept, and this study aims to explore one topic: the level of competence. We employed programming language libraries, functions, and static metrics as proxies for competence, which may not reflect the true theoretical meaning of the concept, and reflect task complexity, PR size, or legacy code structure rather than the contributor's actual capabilities.

\section{Future Work and Conclusion}

Future work will build profiles for AI coding agents as contributors' teammates and will 
explore competence levels with additional metrics, at scale across multiple repositories, applying data source triangulation to assess whether the composite scoring thresholds and predictive models generalize beyond JabRef while exploring diverse LLMs, temperature and tuning parameters. New metric thresholds will be evaluated with industry practitioners using card sort approach. We plan to validate our approach through multiple methods: comparing our competence profiles against self-reported competence and levels in developer profiles, conducting surveys and interviews with project maintainers to assess the relevance of recommended candidates, and conducting experiments to evaluate maintainers' and contributors' feedback about the competences, levels and allocation recommendations. The accuracy to infer training needs, search, and recruitment will be explored in future work, along with an estimation of the costs associated with applying this approach in practice.  Future work will also explore the use of a competence ontology to map the task and contributor profile to perform allocation through ontology matching.
This work presents an initial approach for building competence profiles to later support search, recruitment, and training for the OSS workforce. By implementing competence-based management, we aim to contribute to the sustainability of OSS projects by leveraging complexity level and competence prediction for allocation recommendation.

\section{Data Availability}
The replication package is available through Zenodo \url{https://doi.org/10.5281/zenodo.21605122}. It will be publicly available during the review process and after acceptance.



\bibliography{lipics-v2021-sample-article}
\end{document}